\documentclass[final]{appolb}

\usepackage{amsmath}
\usepackage[usenames,dvipsnames]{xcolor}
\usepackage{graphicx}
\usepackage{dcolumn}
\usepackage{graphicx}

\begin{document}

\title{Towards symmetry-unrestricted Skyrme-HFB in 
       coordinate-space representation: 
       the example of rotational bands of the
       octupole-deformed nucleus $^{222}$Th\thanks{Presented at the XXXV Mazurian Lakes Conference on Physics, Piaski, Poland, September
3-9, 2017}}

\author{W. Ryssens and M. Bender \address{IPNL, Universit\'e de Lyon, Universit\'e Lyon 1, CNRS/IN2P3,\\ F-69622 Villeurbanne, France}\\[5mm]
P.-H. Heenen \address{PNTPM, CP229, Universit\'e Libre de Bruxelles, B-1050 Bruxelles, Belgium}
}

\maketitle

\begin{abstract}
We report on cranked Skyrme-HFB calculations of rotational bands of the 
octupole-deformed nucleus $^{222}$Th. A sudden change in configuration is 
observed, with the shape of the yrast state jumping from large
octupole deformation at low spin to small octupole deformation at high spin.
\end{abstract}

\section{Introduction}

Applications of the self-consistent mean-field approach have a long-standing 
history in nuclear structure physics~\cite{Bender03}. Starting from an energy 
density functional (EDF) that models the effective nucleon-nucleon interaction
in the medium, this  microscopic method provides versatile tools to 
study the structure of nuclei across the entire nuclear chart. 
Their applications are not limited to ground-state properties, but can 
be extended to excited configurations such as $K$-isomers and 
rotational bands.

Recently, we have set up a new code named MOCCa for such calculations 
\cite{RyssensPhd} that is based on the time-tested principles of our 
previous ones~\cite{Bonche05,Bonche87,Bonche1986,Gall1994,Ryssens15a}, 
but unifies them into a single scheme with many new functionalities.
As before, single-particle wave functions are represented on a 3d cartesian 
Lagrange mesh in coordinate space~\cite{Ryssens15a}, and the 
Hartree-Fock-Bogoliubov (HFB) equations are solved with the two-basis
method~\cite{Gall1994}. The code's most important new features concern 
the variational space, where the point-group symmetries \cite{Doba2000a}
imposed in earlier codes can now be individually lifted.

As an example of the large range of physical situations to which the new code 
can be applied, we present results on rotational bands 
of $^{222}$Th. This nucleus has the particularity of having rotational
bands of alternating parity that are connected by strong electric 
dipole transitions~\cite{Smith95}. In a mean-field picture of the nucleus, 
such experimental finding can be associated with a pear-shaped 
octupole-deformed ground-state configuration~\cite{Naza1985,Butler96}.
The study of octupole deformations in this and other regions of the chart 
of nuclei is presently the subject of intense experimental 
activity~\cite{Gaffney2013,Butler16,Reviol14,Bucher16,Maquart17}.

The appearance of octupole-deformed minima in mean-field calculations of the 
deformation energy surface is limited to small regions of the 
nuclear chart around specific combinations of proton and neutron 
numbers~\cite{Robledo11,Ebata17}. With $Z = 90$, some Thorium
isotopes fall into such a region: While reflection-symmetric mean-field minima 
are found for the lighter isotopes up to about $^{218}$Th, calculations for 
those around $^{226}$Th exhibit pronounced octupole-deformed minima, which
again fade away when going to even heavier isotopes. The exact location of the
transition between symmetric and asymmetric shapes depends, however, on 
details of the parameterisation of the EDF used for the calculations. 

Nuclear shapes do not only change with $N$ and $Z$, they also 
can change with angular momentum $J$. Macroscopic-microscopic 
models~\cite{Naza1984,Naza1987} predict that the octupole deformation of 
the ground-state of $^{222}$Th does not persist to high $J$. Direct 
experimental evidence is still lacking, but the observed sudden drop in 
intensity of the populated states at high $J$ could be the fingerprint
of such a change in shape~\cite{Smith95}.

Rotational bands of octupole-deformed nuclei have been studied at high spin  using the cranked HF \cite{Tsvetkov2002} and HFB \cite{Garrote97,Garrote98} methods, 
but, to the best of our knowledge, an abrupt change at high $J$-value has not been put into evidence before.

\section{Skyrme-HFB description of octupole deformed shapes}
\label{sec:ground}
We will first address the description of the ground-state, postponing 
the discussion of rotational bands to the next section. We carry 
out self-consistent HFB calculations using the SLy5s1 parameterisation 
\cite{Jodon16} of the Skyrme EDF in the particle-hole channel. This 
recent fit has been adjusted with a constraint on the surface 
energy coefficient $a_{\rm surf}$ and provides a very satisfying 
description of the deformation properties of heavy nuclei \cite{Ryssens18}. 
For the pairing channel, we add a surface-type contact interaction in
conjunction with the Lipkin-Nogami procedure, with parameters as defined in
Ref.~\cite{Rigollet99}.

As discussed above, both quadrupole and octupole 
correlations play a significant role for nuclei around $^{222}$Th. 
While non-axial degrees of freedom are accessible in our codes, it turns 
out that at intrinsic angular momentum zero the lowest mean-field states 
retain axial symmetry. As a consequence, we can label single-particle 
states by the projection $K$ of their angular momentum on the symmetry 
axis, and characterise nuclear shapes with the dimensionless multipole 
moments $\beta_{20}$ and $\beta_{30}$
\begin{align}
\beta_{20} 
&= \frac{4 \pi}{3 R_0^{2} A} \sqrt{\frac{5}{16 \pi}} \langle 2\hat{z}^2 - \hat{x}^2 - \hat{y}^2 \rangle \, ,\\
\beta_{30} 
&= \frac{4 \pi}{3 R_0^{3} A} \sqrt{\frac{7}{16 \pi}} \langle \hat{z} ( 2\hat{z}^2 - 3\hat{x}^2 - 3\hat{y}^2) \rangle \, ,
\end{align}
where $R_0 \equiv 1.2 A^{1/3} \, \text{fm}$ and the $z$-axis is chosen as reference 
axis. Non-zero values for either indicate broken rotational symmetry in the 
mean-field state, non-zero values for $\beta_{30}$ the additional breaking
of parity. This is accompanied by the loss of parity and angular momentum 
quantum numbers for the many-body as well as the single-particle states. 
As the self-consistent minimisation leads to states for which the
multipole moments $\beta_{\ell m}$ are zero for all combinations of 
odd $\ell$ with $m \neq 0$,
we can, however, impose two spatial point-group symmetries on their 
calculation in a 3d code without influencing the result.
These are $z$-signature $\hat{R}_z$ and the $y$-time simplex 
$\hat{S}^T_y$ as defined in~\cite{Doba2000a}. In addition, for the ground
state of this even-even nucleus we can enforce time-reversal 
symmetry. These are the choices that have been made when setting up 
the Ev4 code for HF+BCS calculations used in~\cite{Bonche1986}.

\begin{figure}
\centering
\includegraphics[width=\linewidth]{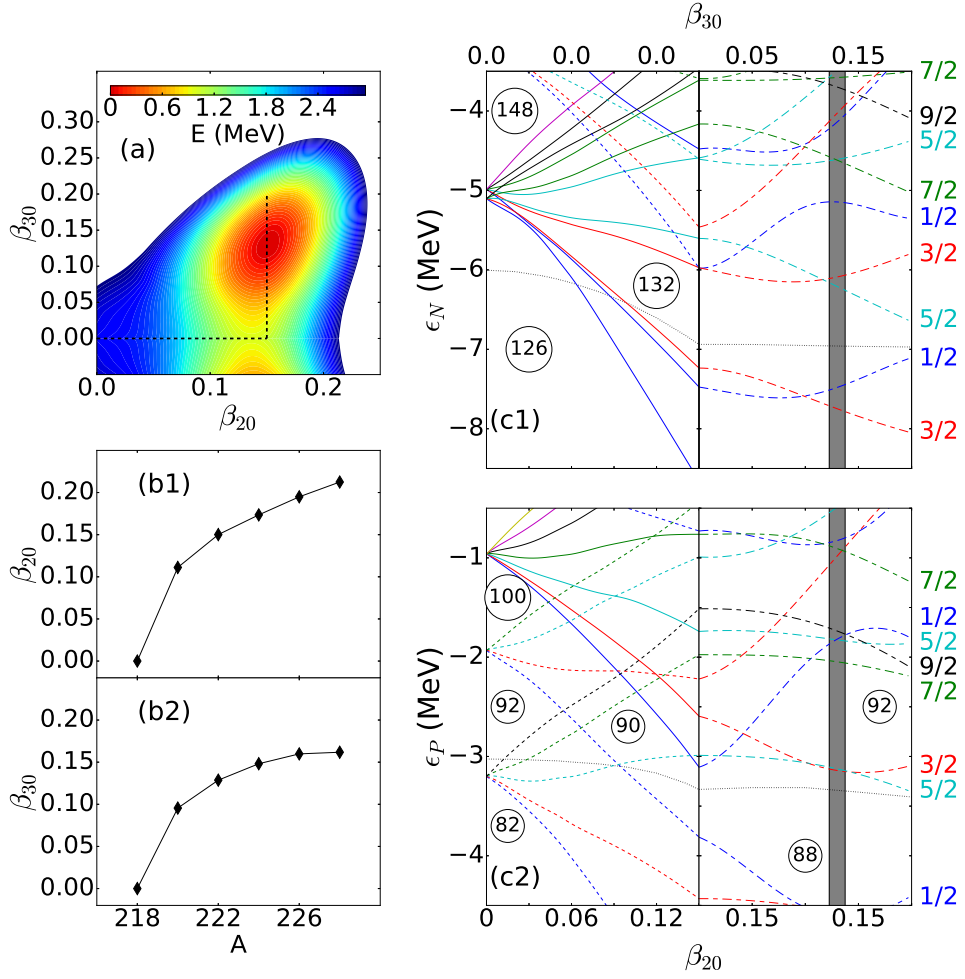}
\caption{
Deformation energy of $^{222}$Th relative to the 
ground state in the \mbox{$\beta_{20}-\beta_{30}$} plane~(a); 
quadrupole~(b1) and octupole deformation~(b2) of the mean-field 
minimum of the Thorium isotopes; Nilsson diagram of neutrons~(c1) 
and protons~(c2) along a path through panel~(a) as indicated by 
the black dashed line. Single-particle states are colour-code according to
their K quantum number. The grey band indicates the location of the overall 
minimum.
}
\label{fig:ground}
\end{figure}

Some results relevant for the analysis of the mean-field energy surface
of $^{222}$Th can be found in 
Fig.~\ref{fig:ground}. Panel~(a) shows the deformation energy as a 
function of $\beta_{20}$ and $\beta_{30}$. A pronounced minimum 
is visible at $(\beta_{20}, \beta_{30}) \approx (0.15, 0.12)$, which is 
about 1~MeV lower than the reflection-symmetric saddle point at $(0.14,0)$.  
This finding differs from relativistic mean-field results obtained with the 
DD-PC1 and NL3 parameterisations~\cite{Nomura2013,Agbemava16}, where it 
is only for $^{224}$Th (DD-PC1) or even $^{226}$Th (NL3) that a minimum 
at comparable deformations appears, while $^{222}$Th remains spherical. 
In fact, in our calculations, static octupole deformation already 
sets in for $^{220}$Th, see panels~(b1) and~(b2) of Fig.~\ref{fig:ground}. 
The same is found for the D1S parameterisation of the Gogny force 
\cite{Robledo11}, whereas for D1M octupole deformation sets in with 
$^{222}$Th \cite{Robledo11}. On the experimental side, rotational
alternating parity bands are observed beginning with $^{222}$Th, while the 
spectrum of $^{220}$Th requires a different interpretation \cite{Reviol14}.

The effect of deforming $^{222}$Th on its single-particle levels is shown in 
panels on the right. They display the Nilsson diagram of eigenvalues of the 
single-particle Hamiltonian for neutrons~(c1) and protons~(c2). For small deformations
up to the vertical  line, these quantities are plotted as a function of 
quadrupole deformation $\beta_{20}$, keeping $\beta_{30} = 0$. Beyond the 
vertical line,
$\beta_{30}$ is varied while $\beta_{20}$ is kept constant at the value of 
the absolute minimum. Different colours indicate the levels' $K$ quantum 
number as listed in the legend. For reflection-symmetric configurations,
$\beta_{30} = 0$, full and short-dashed lines represent levels of positive 
and negative parity, respectively. 
When $\beta_{30} \neq 0$, parity is not a good single-particle quantum number 
and all levels are drawn with long-dashed lines. The Fermi energies of protons
and neutrons are displayed with grey dotted lines.

One can see that the quadrupole and octupole deformations collaborate to 
produce sizeable gaps in both the proton and neutron single-particle spectra 
for $Z=88$, $N=132$ at $(\beta_{20}, \beta_{30}) \approx (0.15,0.12)$. While 
both gaps are also present in the spectra at the symmetric saddle point, 
the octupole deformation results in a further opening of the respective 
gap for both particle species. 
The presence of the same gaps for octupole-deformed nuclei has already 
been reported earlier in Refs.~\cite{Butler96,Engel03}.

\begin{figure}
\center
\includegraphics[width=\linewidth]{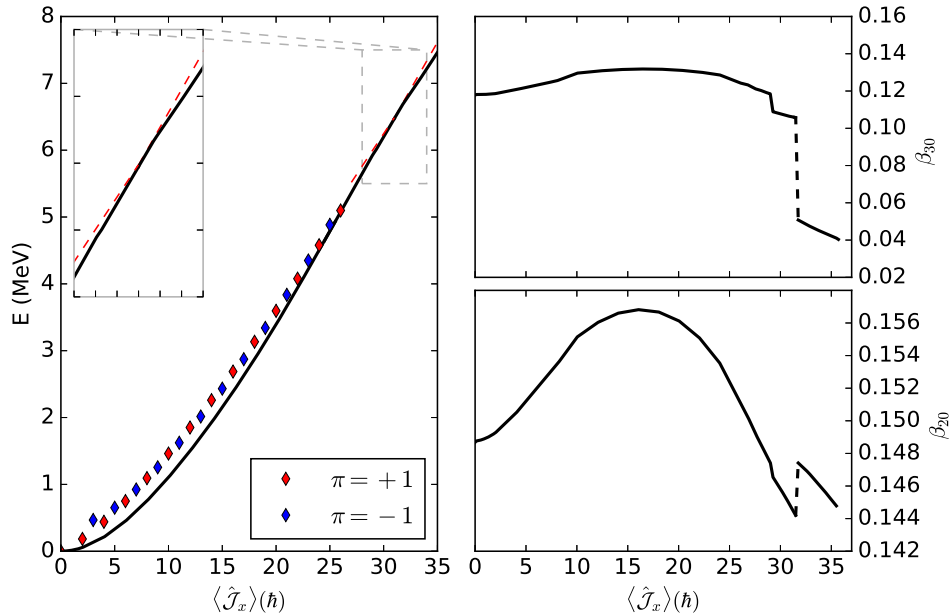}
\caption{
Left: Excitation energy of rotational states of $^{222}$Th as a function of $\langle \hat{\mathcal{J}}_x \rangle$ as obtained in cranked HFB calculations (black solid line) compared to the  experimental data (diamond markers) for yrast states of positive (red) and negative (blue) parity taken from~\cite{Smith95}. 
The inset zooms on the transition region. Dashed lines are linear 
extrapolations of the curve before and after the transition and are 
only shown to guide the eye. The panels on the right display the calculated 
values of $\beta_{30}$ (top) and $\beta_{20}$ (bottom) along the band.}
\label{fig:JXE}
\end{figure}

\section{Cranked Skyrme-HFB for rotating octupole shapes}
\label{sec:crank}

To describe rotational bands in a self-consistent mean-field approach, 
it is customary to vary the energy subject to a cranking constraint 
\cite{Bonche87,Satula05}
\begin{equation}
\label{eq:routhian}
R = E - \text{\boldmath$\omega$} \cdot \langle \hat{\text{\boldmath$\mathcal{J}$}} \rangle \, .
\end{equation}
While our new code can handle arbitrary orientations of the many-body 
angular momentum vector {\boldmath{$\mathcal{J}$}}, we will limit ourselves 
here to rotation about a principal axis perpendicular to the symmetry axis
of the ground state of $^{222}$Th, chosen to be the $x$-axis. 
The presence of the constraint in Eq.~\eqref{eq:routhian} requires one
to lift several of the symmetries that could be imposed on calculations
of that nucleus' ground state. First, $\hat{\mathcal{J}}_x$ is a 
time-odd operator, such that time-reversal symmetry has to be abandoned.
Second, finite values of $\beta_{30}$ are only compatible with 
conserved $z$ signature, while non-zero expectation values of 
$\langle \hat{\mathcal{J}}_x \rangle$ are only compatible with $x$ signature.
Their combination rules out any conserved signature symmetry~\cite{Doba2000a}.
When pointing into a direction that is not a symmetry axis of the nucleus,
finite values of $\langle \hat{\mathcal{J}}_x \rangle$ are also 
incompatible with axial symmetry, such that the solution of the cranked
HFB equations necessarily requires a 3d code as used here. From the 
symmetries mentioned in Sect.~\ref{sec:ground}, only $\hat{S}^T_y$ has been
retained. All converged many-body states discussed here, however, conserve 
$x$-simplex $\hat{S}_x$ as defined in~\cite{Doba2000a}, although it had not 
been imposed.
The calculations can be carried out either for constant rotational 
frequency $\omega_x$ or such that the solution takes a specific value for 
$\langle \hat{\mathcal{J}}_x \rangle$. Both has been used to construct
states in the yrast band for which the energy, the quadrupole and the 
octupole deformation are shown in Fig.~\ref{fig:JXE}. At low spin, the 
behaviour of the energy is quadratical as it would be the case for a deformed 
classical rotor. At intermediate spins, deformations change slowly and the 
response of the energy to the cranking constraint becomes closer 
to linear. Between $\langle \hat{\mathcal{J}}_x \rangle = 31$ and $32$, 
however, the calculated energy shows a kink indicating an abrupt structural 
change in the system.  At this point, all deformations exhibit a 
discontinuity. The change is very pronounced for the octupole deformation, 
decreasing from a large value $\beta_{30} \approx 0.12$ to a much smaller 
one of $\beta_{30} \approx 0.05$. In comparison, $\beta_{20}$
changes on a much smaller scale, which is a consequence of the rigidity of the 
potential energy surface in the quadrupole direction. A similar result 
has been obtained in a microscopic-macroscopic model~\cite{Naza1984,Naza1987},
but already at an angular momentum of about $24$. 

The calculated energies do not exhibit exactly the same behaviour as the 
available data. As cranked HFB is designed to describe high-spin states, some
disagreement at low spin is not unexpected. It also has to be noted that 
pure cranked HFB can only calculate the energies of some mixture of 
states in the positive- and negative-parity bands of an octupole-deformed
nucleus \cite{Garrote97,Garrote98}. As at all spins the energy surface is 
quite soft in octupole  direction, any small change in the effective 
interaction might have a visible impact on the moment of inertia of this
nucleus.

\section{Summary and Outlook}
\label{sec:conclusion}

We have studied the evolution of the shape of yrast states in the octupole
deformed $^{222}$Th as a function of spin in cranked HFB calculations. 
These calculations are technically challenging, as they require a 3d code 
in which the often imposed time-reversal, parity and signature symmetries 
are simultaneously lifted. The loss of the latter significantly complicates 
the numerical solution of the HFB equations~\cite{RyssensPhd}.

Our calculations predict that yrast states undergo a sudden change from 
shapes with large octupole deformation to almost symmetric ones
when spinning up the nucleus. Such transition has not been
observed yet, but might occur for states at higher spins than those populated 
in past experiments~\cite{Smith95}. New experimental data for the odd-mass 
neighbour $^{223}$Th point indeed towards such a shape change \cite{Maquart17}. 
A detailed discussion of this system along the lines of the present study is in
preparation.

\section{Acknowledgments}

The computations were performed using HPC resources from the 
computing centre of the IN2P3/CNRS and the Consortium des {\'E}quipements
de Calcul Intensif (C{\'E}CI), funded by the Fonds de la Recherche 
Scientifique de Belgique (F.R.S.-FNRS) under Grant No.~2.5020.11.
W.R. and P.-H.H. gratefully acknowledge funding by the IAP Belgian Science 
Policy (BriX network P7/12).

\end{document}